\documentclass[apj, iop]{emulateapj}
    \usepackage{amsmath}
    \usepackage{color}

\shorttitle{A Catastrophe Model for Coronal Mass Ejections}
\shortauthors{Wang et al.}

\begin{document}

\title{Contribution of Velocity Vortices and Fast Shock Reflection and Refraction to the Formation of EUV Waves in Solar Eruptions}

\author{Hongjuan Wang\altaffilmark{1}, Siqing Liu\altaffilmark{1},
 Jiancun Gong \altaffilmark{1}, Ning Wu \altaffilmark{2} and Jun Lin \altaffilmark{3}}

\altaffiltext{1}{Center for Space Science and Applied Research, Chinese Academy of Sciences,
Beijing 100190, China}
\altaffiltext{2}{School of Tourism and Geography, Yunnan Normal
                University, Kunming, Yunnan 650031, China}
\altaffiltext{3}{Yunnan Observatories, Chinese Academy of
              Sciences, Kunming, Yunnan 650011, China}

\begin{abstract}
We numerically study the detailed evolutionary features
of the wave-like disturbance and its propagation in the eruption. 
This work is a follow-up to Wang et al., using significantly upgraded new simulations.
We focus on the contribution of the velocity vortices and 
the fast shock reflection and refraction in the solar corona to the formation of the EUV waves.
Following the loss of equilibrium in the coronal magnetic structure,
the flux rope exhibits rapid motions and
invokes the fast-mode shock forward of the rope, which then produces the
type II radio burst. The expansion of the fast shock, 
which is associated with outward motion, takes place in various directions, and the downward expansion shows the reflection and
the refraction as a result of the non-uniform background plasma.
The reflected component of
the fast shock propagates upward and the refracted component propagates
downward. As the refracted component reaches the boundary
surface, a weak echo is excited.
The Moreton wave is invoked as the fast shock touches the bottom boundary, so the Moreton wave
lags the type II burst. A secondary echo occurs in the area where reflection of the fast shock
encounters the slow-mode shock, and
the nearby magnetic field lines are further distorted because of the interaction between
the secondary echo and the velocity vortices.
Our results indicate that the EUV wave may arise from various processes that are
revealed in the new simulations.
\end{abstract}


\section{Introduction}

The most intense energetic activity in the solar system is the solar eruption
that produces solar flare, eruptive prominence, and coronal mass ejection (CME).
During the process, a large number of magnetized energetic plasmas (with mass of up to
$10^{16}$ g and energy of $10^{32}$ erg) are ejected into the
interplanetary space within a short timescale, and hence disturb
spatial and planetary magnetic field and significantly affect
satellite operation, aviation power, human space exploration,
communication and so on (\citet{Schwenn06,Pulkkinen07,Lin07,Chen11a,Cheng12}
and references therein). One interesting phenomenon closely associated
with CMEs is the globally propagating wave-like disturbance in the
corona, i.e. the EIT or EUV wave. \citet{Moses97} first reported this
phenomenon, and \citet{Thompson98} first
analyzed this phenomenon in detail using the data from the
Extreme-ultraviolet (EUV) Imaging Telescope (EIT) on (space) board the
\emph{Solar and Heliosphere Observatory} (SOHO) spacecraft.

The EUV waves have generally been observed as broad, diffuse arc-shaped
bright front with lifetime of about 50 minutes. It could be seen in the lower
corona (at temperature of 1-2 MK). The speed of the EUV wave front varied
from about 50 to over 700 km~s$^{-1}$ with `typical' speed of 200-400
km~s$^{-1}$ \citep{Thompson09}. The EUV waves are usually associated
with CMEs, dimmings, type II radio bursts, and flares \citep{Biesecker02}.
Recently, \citet{Nitta13} presented a large sample of events that
look like EUV waves as observed by the Atmospheric Imaging Assembly (AIA;
\citet{Lemen12} on board the \emph{Solar Dynamics Observatory} (\emph{SDO}),
and revisited their associations with flares, CMEs and type II radio bursts.
They found that the speed of EUV waves is not strongly correlated with CME
magnitude or the flare intensity, nor do they show an association with type
II bursts (cf. \citet{Nitta14}).

Observations so far have shown that the EUV waves may largely be explained in terms of
coronal fast-mode magnetohydrodynamic (MHD) wave (e.g., \citet{Thompson99,Warmuth04a,
Warmuth04b,Gopalswamy09,Patsourakos09,Schmidt10,Long11,Liu11,Zheng11,Zheng12,Zheng13a,
Zheng14,Cheng12,Shen12b,Shen12a,Yang13,Shen14b}) driven by CME-related eruptions
although other components may co-exist in the same event or exist in other events
\citep{Patsourakos12}. It is possible to identify the properties of EUV waves
that can be naturally linked with fast-mode shock waves. Observations showed that in
their early stages EUV waves either experience significant deceleration \citep{Cheng12,Shen12b,Yang13}
or propagate at approximately constant speeds \citep{Ma09,Patsourakos09,Liu10,Long11}.
\citet{Warmuth11} and \citet{Nitta13} investigated both cases on the basis of a large number of samples.

Regardless of their initial speeds or deceleration profiles, these waves
end up travelling within a speed range of 180-380 km s$^{-1}$, which is
consistent with the fast-mode speed over the quiet Sun \citep{Downs11,Patsourakos12}.

In addition, two other characteristics of EUV waves are found in
case studies. First, what appears to be a single wave may in fact consist of fast and
slow components. In the 2010 April 8 event, \citet{Liu10} identified the
two components and explained the large-scale ripples in terms of the faster
component overtaking the slower one. Second, EUV waves are seen to reflect
and refract as they propagate in the corona. See, for example,
\citet{Shen13}, who studied the 2012 April 23 event. Furthermore, \citet{Yang13} noticed in
the 2011 August 4 event that a secondary wave was excited by a reflected wave.

Although the fast-mode MHD wave scenario of the EUV waves is supported by
a lot of observational and theoretical results
\citep{Thompson99,Warmuth04a,Warmuth04b,Gopalswamy09,Patsourakos09,Schmidt10,Long11,Liu11,Li12,
Zheng11,Zheng12,Cheng12,Shen12b,Shen12a,Yang13}, there is little understanding
of how it evolves once the fast-mode shock is produced by solar eruptions.
Therefore, numerical experiments are needed to better understand
the details of how the EUV waves evolve. We demonstrate it here as our main goal of this work.

In this paper, we numerically study the detailed wave-like disturbance
and its propagation to uncover processes not commonly discussed but potentially
important to understand some aspects of the EUV waves. In our simulations, the
employment of the high resolution grid and the empirical atmosphere model
(\citet{Sittler99}, hereafter S\&G) provides a good opportunity to study
some important features that were not shown in our previous work \citep{Wang09,Mei12}.
We describe the physical model, formulae and numerical approaches in the next section.
In Section 3, the numerical results are presented, and we discuss these results
in Section 4. Finally, we summarize this work in Section 5.

\section{Numerical Model and Formulae}

We suppose a two-dimensional magnetic configuration in the
semi-infinite $x$-$y$ plane. In the coordinates, $y=0$ corresponds
to the boundary between the photosphere and the chromosphere. The
chromosphere and the corona are represented by $y>0$. In this model,
we use a current-carrying flux rope to represent
a prominence or filament that floats in the corona, and a line dipole below the photosphere to
denote the photospheric background field. The evolution of the magnetic system satisfies the following
ideal MHD equations:

\begin{eqnarray}
\begin{split}
&\frac{\textrm{D}\rho}{\textrm{D}t}+\rho\nabla\cdot \textbf{v}=0,
\label{eq:cont}\\
&\rho\frac{\textrm{D}\textbf{v}}{\textrm{D}t}=-\nabla
p+\frac{1}{c}\textbf{J}\times \textbf{B}+\rho\frac{GM_{\odot}}{(R_{\odot}+y)^{2}}, \label{eq:momen}
\\
&\rho\frac{\textrm{D}}{\textrm{D}t}(e/\rho)=-p\nabla\cdot
\textbf{v},\label{eq:energy}\\
&\frac{\partial \textbf{B}}{\partial t}=\nabla\times(\textbf{v}\times
\textbf{B}),\label{eq:induc}\\
&\textbf{J}=\frac{c}{4\pi}\nabla\times \textbf{B},\label{eq:current}\\
&p=(\gamma-1)e,\label{eq:p1}\\
&p=\rho kT/m_{p}, \label{eq:p2}
\end{split}
\end{eqnarray}
where $\textbf{B}$ indicates the magnetic field, $\textbf{J}$ the electric current density,
$\textbf{v}$ the velocity of the flow, $\rho$ the mass density, $p$ the gas pressure, $e$
the internal energy density, $G$ the gravitational constant, $M_{\odot}$ the solar mass,
$R_{\odot}$ the solar radius, $\gamma$ the ratio of specific heats, $m_{p}$ the proton mass.
The ZEUS-2D MHD code \citep{Stone92a,Stone92b,Stone92c} is employed to solve Equations (1).

In our simulations, the magnetic configuration is composed of the current-carrying flux rope,
the image of the current inside the flux rope, and the background magnetic field that is generated
by a line dipole located at $y=-d$ below $y=0$. The relative strength of the dipole field $M$ is denoted
by a dimensionless parameter $M=m/(Id)$, which is related to the ratio of the strength of the
dipole field $m$ and the product of the filament current $I$ and the depth $d$ of the dipole
field \citep{Forbes90,Wang09}. The flux rope is located at $y=h$ above the bottom boundary.

The initial magnetic configuration from which the eruption starts is:

\begin{eqnarray}
B_{x}&=&B_{\phi}(R_{-})(y-h_{0})/R_{-}-B_{\phi}(R_{+})(y+h_{0})/R_{+}\nonumber\\
&-&B_{\phi}(r+\Delta/2)Md(r+\Delta/2)[x^{2}-(y+d)^{2}]/R^{4}_{d},
\label{eq:Bx}\\
B_{y}&=&-B_{\phi}(R_{-})x/R_{-}+B_{\phi}(R_{+})x/R_{+}\nonumber\\
&-&B_{\phi}(r+\Delta/2)Md(r+\Delta/2)2x(y+d)/R^{4}_{d},\label{eq:By}
\end{eqnarray}
with
\begin{eqnarray*}
R^{2}_{\pm}&=&x^{2}+(y\pm h_{0})^{2},\\
R^{2}_{d}&=&x^{2}+(y+d)^{2},
\end{eqnarray*}
and $B_{\phi}(R)$ is determined by the electric current density distribution $j(R)$
inside the flux rope. They are:
\begin{equation}
\small
\begin{split}
&B_{\phi}(R) = -\frac{2\pi}{c}j_{0}R, ~~ 0\leq R\leq r-\Delta/2;\\
&B_{\phi}(R) = -\frac{2\pi j_{0}}{c R}\Bigg\{\frac{1}{2} \Big(r -
\frac{\Delta}{2}\Big)^{2} - \Big(\frac{\Delta}{\pi}\Big)^{2} +
\frac{1}{2}R^{2} \\
&+ \frac{\Delta R}{\pi}
\sin\Big[\frac{\pi}{\Delta}\Big(R-r+\frac{\Delta}{2}\Big)\Big]\\
&\Big.+ \Big. \Big(\frac{\Delta}{\pi}\Big)^{2}
\cos\Big[\frac{\pi}{\Delta}\Big(R-r+\frac{\Delta}{2}\Big)\Big]\Bigg\},
 r- \Delta/2 < R <r + \Delta/2;\\
&B_{\phi}(R) = -\frac{2\pi j_{0}}{c
R}\left[r^{2}+(\Delta/2)^{2}-2(\Delta/\pi)^{2}\right], r+\Delta/2\leq R<\infty;\\
&j(R) = j_{0},  0\leq R\leq r-\Delta/2;\\
&j(R) = \frac{j_{0}}{2}{\cos[\pi(R-r+\Delta/2)/\Delta]+1}, r-\Delta/2<R<r+\Delta/2,\\
&j(R) = 0,  r+\Delta/2\leq R<\infty.
\end{split}
\end{equation}

The initial background plasma density $\rho_{0}(y)$ in this work
is based on the S\&G model, and therefore more realistic than that
used in the previous model \citep{Wang09,Mei12}. We have:
\begin{eqnarray}
\small
\begin{split}
&\rho_{0}(y)=\rho_{00}f(y),\\
&f(y)=a_{1}z^{2}(y)e^{a_{2}z(y)}\Big[1+a_{3}z(y)+a_{4}z^{2}(y)+a_{5}z^{3}(y)\Big],\\
&z(y)=\frac{R_{\odot}}{R_{\odot}+y}. \label{eq:rho0}
\end{split}
\end{eqnarray}
Here we adopt $\rho_{00}=1.672\times 10^{-13}$ g~cm$^{-3}$,
which is about one order of magnitude smaller than that in our previous
work \citep{Wang09}, and $y$ denotes the height from the surface of the Sun.
We take $a_{1}=0.001292$, $a_{2}=4.8039$, $a_{3}=0.29696$, $a_{4}=-7.1743$, $a_{5}=12.321$.

The S\&G model, which is two-dimensional and semi-empirical,
was developed using the {\it Skylab} white-light coronagraph
\citep{Guhathakurta96} and {\it Ulysses} in-situ \citep{Phillips95} measurements.
It smoothly connects the density distributions near and
far from the Sun (see Figure 1 of \citet{Lin02}). The density distribution $f (y )$ given
by Equation (5) describes an isothermal atmosphere. Density decreases exponentially with height
in the lower corona, and then decreases much more slowly as $y^{-2}$.
The results of radio observations of type III bursts over a wide frequency band
from a few kHz to 13.8 MHz also suggested the $y^{-2}$  variation of the plasma density
far from the Sun \citep{Leblanc98}. This means that the atmosphere model with density distribution
given in Equation (5) could be considered as a realistic model. In addition,
we choose to use the present values of coefficients for the density profile that describes
the density distribution in the polar region.

\begin{figure}
\centering
\includegraphics[width=9cm,clip,angle=0]{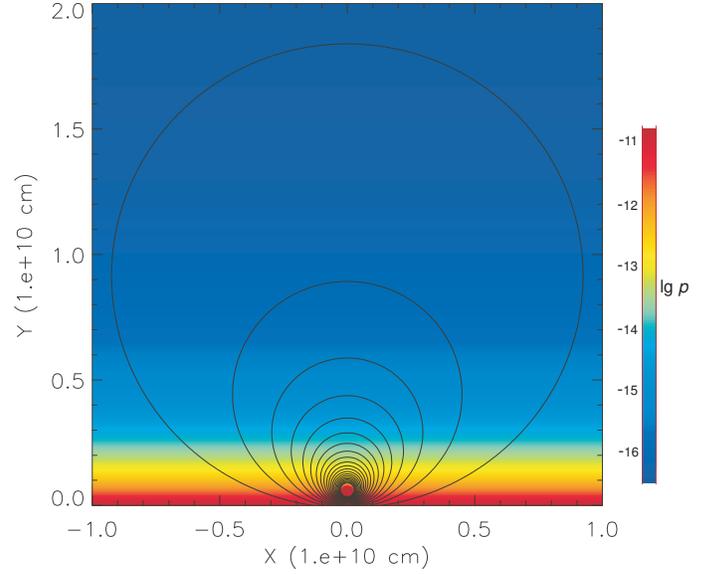}
\caption{Distributions of the initial configuration of the magnetic
field (black contours) and the plasma density (shadings).
The right color bar represents values of the density in $\lg\rho$~(g~cm$^{-3})$.}
\label{fig:1}
\end{figure}

S\&G also gives another profile for the density distribution in the equatorial
region where the helmet streamer is located (see equation [18a] and Table 1
of S\&G's original paper). Although the coefficients for these two profiles
are different, the values of the resultant density distributions in the two cases
are not apparently different (see Figure 2 of S\&G). This means
that using either profile in our calculations would not affect the final results and conclusions in a major way.
Figure \ref{fig:1} depicts the distributions of the initial
configuration of the magnetic field (black contours) and the plasma density (shadings)
used in the present work.

We need to note here that the plasma density on the coronal base
($y = 0.0$) adopted in this work is about an order of magnitude
higher than that in reality, but the magnetic field strength
is roughly comparable to what is usually assumed. As a result,
the Alfv\'{e}n and related speeds in our work are about a factor
of three lower than those that commonly appear in the literature.
A lower Alfv\'{e}n speed allows the fast shock to form more easily
since we in this work aim to study the response of various
layers of the solar atmosphere to the fast-mode shock driven by
the lift-off of the flux rope. This may cause slower propagation
and weaker strength of the shock, but will not change other properties of
the shock.

There is a balance between pressure gradient of gas and the gravity for the initial background atmosphere as below:
\begin{eqnarray}
\nabla p_{0}(y)=-\rho_{0}(y)\frac{GM_{\odot}}{(R_{\odot}+y)^{2}}. \label{eq:p0}
\end{eqnarray}
Taking Equations (\ref{eq:rho0}) into (\ref{eq:p0}), we can get the initial background pressure $p_{0}(y)$.
Then the temperature
distribution $T_{0}(y)$ is obtained as follows
\begin{eqnarray}
p_{0}(y)=\frac{\rho_{0}(y)}{m_{p}}kT_{0}(y), \label{eq:T0}
\end{eqnarray}
where $k$ is the Boltzmann constant.

The initial total pressure consists of the gas pressure and the
magnetic pressure. The initial total pressure and the mass density can be written as
\begin{eqnarray}
\begin{split}
&p=p_{0}-\int^{\infty}_{R_{-}}B_{\phi}(R)j(R)dR,\\
&\rho=\rho_{0}(p/p_{0})^{1/\gamma}. \label{eq:p3}
\end{split}
\end{eqnarray}

The computational domain is $(-4L, 4L)\times (0, 8L)$ with
$L=10^{5}$ km, and the grid points is $800\times 800$.
The bottom boundary at $y=0$ applies a line-tied condition, while
the other three use the open boundary. Table \ref{tbl:1} lists the initial values of the
parameters in our simulation.

\begin{table*}
\caption{Initial values for several important parameters of the
numerical experiment}
\center
\begin{tabular}{lll}
\hline
$r_{0}=2.5\times 10^{3}$ km & $h_{0}=6.25\times 10^{3}$ km  & \\
$\rho_{00}=1.672\times10^{-13}$ g~cm$^{-3}$ & $T_{00}=10^{6}$ K & $j_{00}=1200$ statamp~cm$^{-2}$ \\
\hline
\end{tabular}
\label{tbl:1}
\end{table*}

\section{Results of Numerical Experiments}

In this section, we present the results of our numerical experiments.
When the magnetic compression in our experiments surpasses
the magnetic tension, the flux rope quickly rises from the beginning.
Some important concomitant phenomena appear as the flux rope rises,
such as the fast-mode shock around the flux rope, the slow-mode
shock emerging from two sides of the flux rope, and the velocity vortices.
In order to let us more easily find the region where the fast-mode shock occurs,
we plot the distributions of the fast-mode magneto-acoustic
speed, $c_{f}$, at time $t=0$~s and $t=50$~s, respectively, as the eruption is in progress.
We here have $c_{f}^{2}=v_{A}^{2}+c_{s}^{2}$ with $v_{A}=|B|/\sqrt{4\pi\rho}$ and
$c_{s}=\sqrt{\gamma p/\rho}$ being the Alfv\'{e}n speed and sound speed, respectively.

\begin{figure}
\centering
\includegraphics[width=9cm,clip,angle=0]{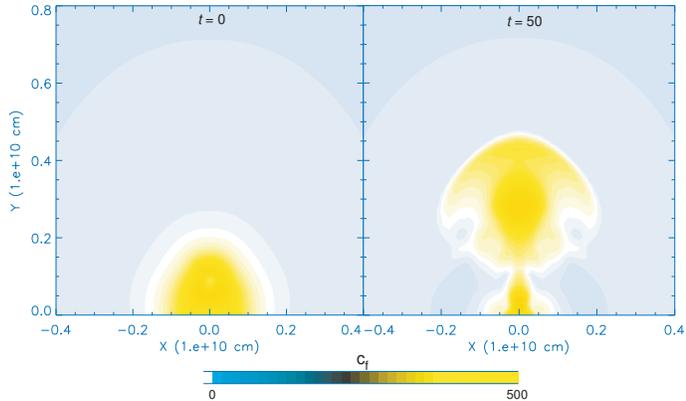}
\caption{The fast-mode speed $c_{f}$ contours at $t = 0$~s and $t = 50$~s. The lower color bar represents values of the speed in km s$^{-1}$.}
\label{fig:2}
\end{figure}

We studied disturbances caused by the interaction
of the fast-mode shock with the boundary to explain the Moreton wave,
and interaction between the slow shock and the velocity vortices to
account for the EUV waves \citep{Wang09}. In this work, we focus on the formation and
propagation of the reflection/refraction of the fast-mode shock and
the velocity vortices to investigate their contribution to the formation
of the EUV waves.

\citet{Mei12} found that in the lower Alfv\'{e}n speed environment,
the velocity vortices and the slow shock may be responsible for EUV waves.
In the higher Alfv\'{e}n speed environment, the second
echo of the fast-mode shock together with the slow shock and the
velocity vortices may produce EUV waves. However, our results show
that even in the same Alfv\'{e}n speed environment, various origins
of waves could be easily recognized due to the very high grid resolution
used in our simulations.

\subsection{Formation of the vortices and the fast shock}

\begin{figure}
\centering
\includegraphics[width=9cm,clip,angle=0]{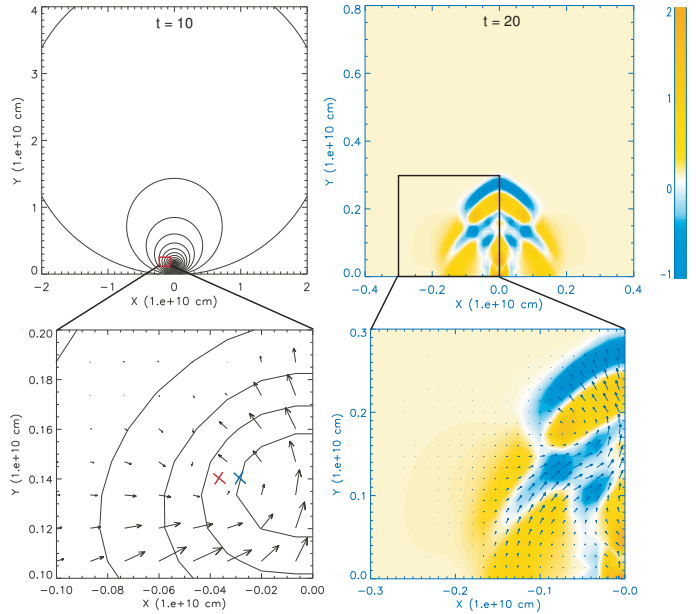}
\caption{The velocity divergence $\nabla\cdot \mathbf{v}$ and streamlines at $t = 10$~s and $t = 20$~s.
The lower panels provide detailed structures in two subregions surrounded by rectangles marked in the upper panels,
respectively. The blue cross shows the location of the left edge of the flux rope, and the red cross denotes
the location of the velocity vortices at 10~s. The unit of time is second. The right color bar
represents values of the velocity divergence in arbitrary unit.}
\label{fig:3}
\end{figure}

The loss of equilibrium results in fast upward movement of the flux rope
and formation of a region with lower pressure behind the flux rope
\citep{Forbes90,Wang09}.
This causes two flows. They are
toward the lower pressure region, fetching in both magnetic fields of opposite
polarity and plasma. Forcing magnetic fields of opposite to flow together
consequently leads to magnetic reconnection in this region. The velocity
vortices on either side of the reconnection region forms as shown in Fig.\,\ref{fig:3}.

\begin{figure}
\centering
\includegraphics[width=9cm,clip,angle=0]{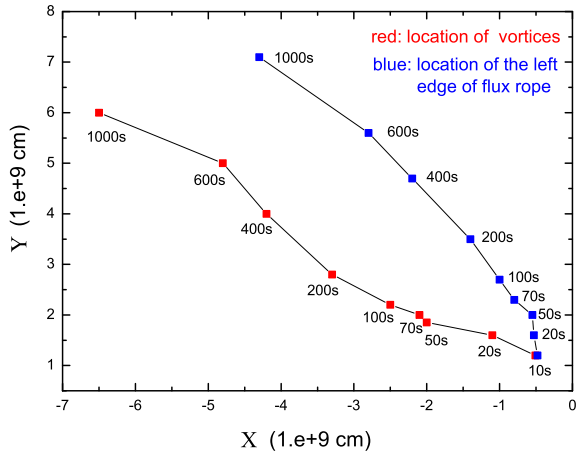}
\caption{The location of the velocity vortices on the left side of the flux rope, and the location
of the left edge of the flux rope at various times.}
\label{fig:4}
\end{figure}

To look into more details of the velocity vortices and their impact on the flux rope,
we enlarged the region of interest and created a composite of the velocity divergence
$\nabla\cdot \mathbf{v}$ and streamlines in Fig.\,\ref{fig:3}.
In the lower left panel, the blue and red crosses show, respectively, the locations of
the left edge of the flux rope and the velocity vortices at 10~s, respectively.
Figure \ref{fig:4} shows the locations of the velocity vortices on the left side of
the flux rope (in red) and the left edge of the flux rope (in blue) at various times.

From Figs.\,\ref{fig:3} and \ref{fig:4}, we notice that the velocity vortices
blend with the edge of the flux rope at $t = 10$~s. It is not easy to make a distinction
between the velocity vortices and the edge of the flux rope.
However, they start to separate, and the distance between them increases with time.

\begin{figure}
\centering
\includegraphics[width=9cm,clip,angle=0]{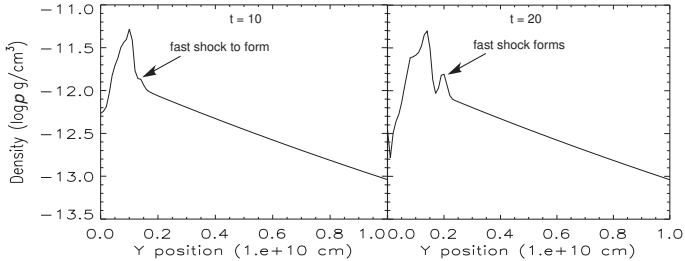}
\caption{Plasma density distribution along the y-axis at $t = 10$~s and $t = 20$~s. The highest peak in two
panels represents the flux rope. In the panel of $t = 10$~s, a slight increase indicates the formation of the
fast-mode shock, and in the panel of $t = 20$~s, the fast shock is formed.}
\label{fig:5}
\end{figure}

Another important finding is that a fast-mode shock starts to form
in front of the flux rope at the time when the velocity vortices almost separate from
the flux rope. To reveal more details of this process,
we check the formation time of the fast-mode shock as shown in Fig.\,\ref{fig:5}.
It plots variations of the plasma density versus height along the $y$-axis
at two different times. The highest peaks in both panels indicate the flux rope.
In the panel of $t = 10$~s, we can recognize a slight increase in the density
right forward of the flux rope, which represents the formation of the fast-mode shock.
The fast shock is more clearly seen in the panel of t = 20 s, which may correspond
to the occurrence of type II radio bursts as discussed by
\citet{Wang09} and \citet{Lin06}. Just during this period,
the velocity vortices begin to separate from the edge of the rope.

\begin{figure}
\centering
\includegraphics[width=9cm,clip,angle=0]{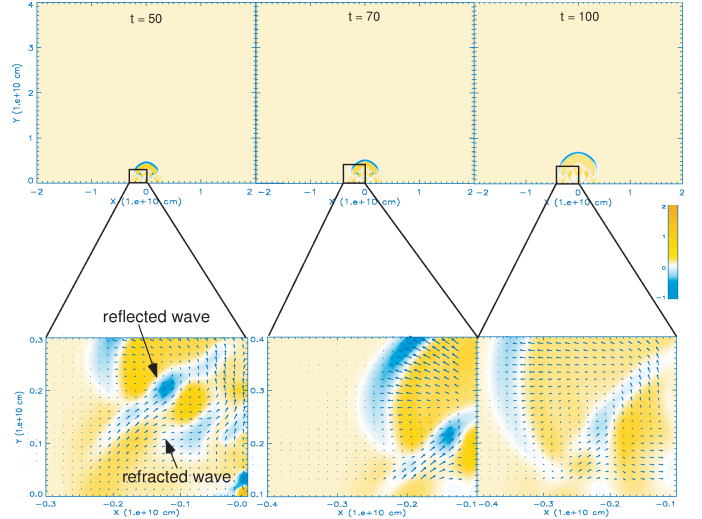}
\caption{Evolutions of the velocity divergence $\nabla\cdot \mathbf{v}$ and streamlines at various times.
The lower panels provide detailed structures in three subregions surrounded by rectangles marked in the upper panels,
respectively. The right color bar represents values of the velocity divergence in arbitrary unit.}
\label{fig:6}
\end{figure}

\begin{figure}
\centering
\includegraphics[width=9cm,clip,angle=0]{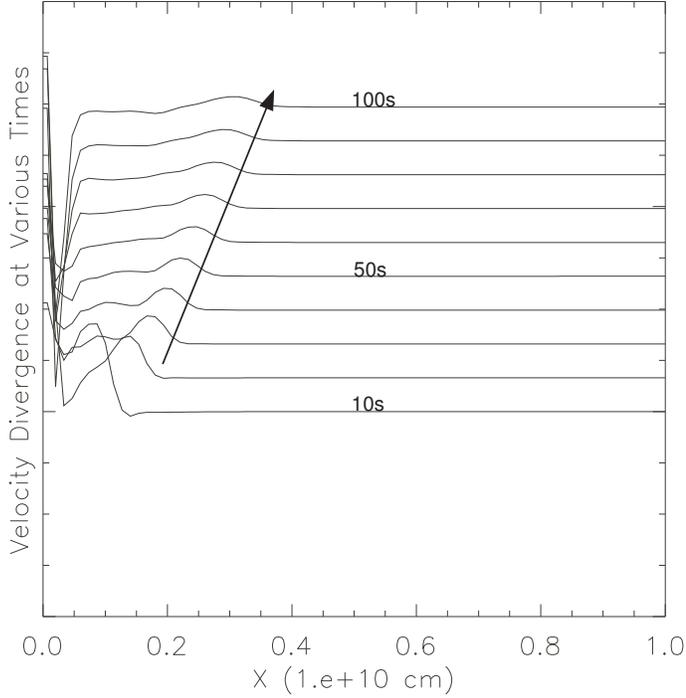}
\caption{Evolution of $\nabla\cdot \mathbf{v}$ on the layer $y = 1.4\times10^{4}$~km at various times.}
\label{fig:7}
\end{figure}

Figure \ref{fig:6} plots the evolutions of $\nabla\cdot \mathbf{v}$ and streamlines
as the flux rope moves outward. Three regions surrounded
by rectangles display places where the flow is strong.
Details in these three regions can be seen more clearly
in the lower enlarged panels in Fig.\,\ref{fig:6}.
The plasma flow manifests apparent vortices near the side
back of the flux rope. In each panel, a crescent feature
around the flux rope represents the fast-mode shock.
With the fast shock propagating forward, it expands sideward
and backward, and its footprints touch the regions
of the vortices at about $t = 20$~s. To obtain
more details of this process, we studied the behavior
of the distribution of $\nabla\cdot \mathbf{v}$ on the
layer of the vortices, i.e. $y = 1.4\times10^{4}$~km.
Ten curves in Fig.\,\ref{fig:7} are for the distributions
of $\nabla\cdot \mathbf{v}$ on this layer ($y = 1.4\times10^{4}$~km)
at different times.
Plots in Fig.\,\ref{fig:7} clearly show the fast shock that sweeps
the layer $y = 1.4\times10^{4}$~km from $t = 20$~s,
but there is no sign of the fast-mode shock before $t = 10$~s.
The corresponding speed of the fast shock in this layer is about 300~km~s$^{-1}$.
Figure \ref{fig:3} suggests that the
footprint of the fast shock does not arrive at the layer where the
vortices occur until $t = 20$~s. Our simulations reveal that the impact of
the fast shock joins that of the velocity
vortices from $t = 20$~s, inducing more complex patterns of flow or
disturbance in this layer. On the other hand, the fast-mode shock itself
does not stay in this layer. Instead, it continues to expand outward,
leaving behind the disturbances that it causes in the corona.
Obviously, the fast shock propagates in the corona faster than the
disturbances that it produces. This might account for the fast component
of EUV waves usually observed in the same coronal layer.

\subsection{Formation and propagation of the fast shock reflection and refraction}

\begin{figure}
\centering
\includegraphics[width=9cm,clip,angle=0]{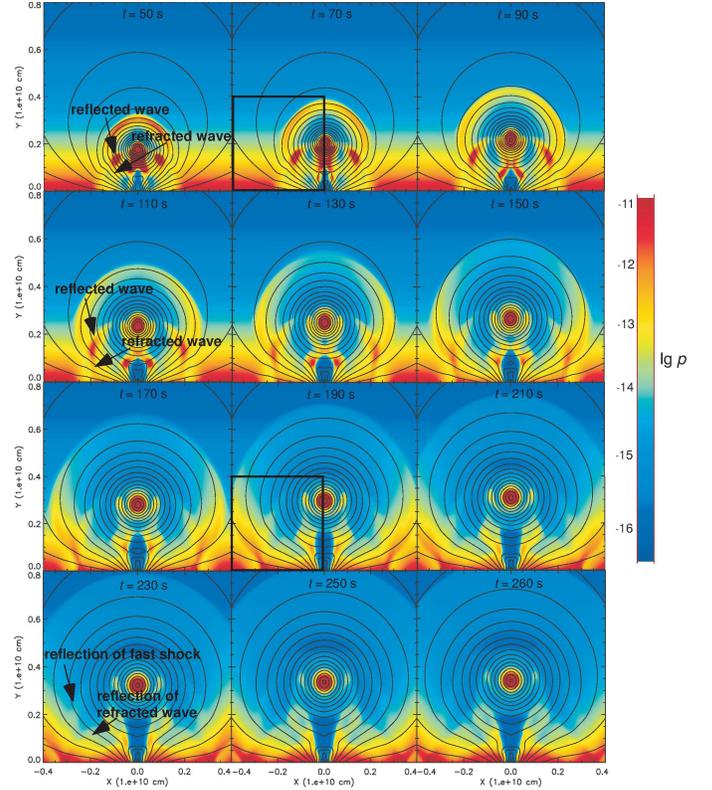}
\caption{Set of snapshots of magnetic field (black contours) and
the plasma density (shadings) interval between 50 s and 260 s.
The right color bar represents values of the density in $\lg\rho$~(g~cm$^{-3})$.
Two regions surrounded by rectangles at $t = 70$~s and $t = 190$~s display areas where
the angle of the reflection and the angle of the refraction vary with the incident angle
growing large. Details in these two regions can be seen more clearly in Figure \ref{fig:9}.}
\label{fig:8}
\end{figure}

In addition to the velocity vortices and the fast-mode shock,
one more feature is shown in Figs.\,\ref{fig:6} and \ref{fig:8}.
A composite of the plasma density (shadings) and
magnetic field lines (black contours) during the interval
between 50 s and 260 s can be seen in Fig.\,\ref{fig:8}.
At t=50 s, the plasma flow near the side back of the flux rope results in vortices,
as clearly seen in the lower-left panel of Fig.\,\ref{fig:6}.
The flow is so strong that it impacts the nearby magnetic field lines,
resulting in distinct deformation of
the nearby magnetic structures. Because the background plasma in the corona is not uniform, reflection and
refraction take place at both wake ends of
the fast shock as it moves downward (see Fig.\,\ref{fig:8}).
The reflected component of the fast shock propagates upward and
the refracted component propagates downward continuously.
In this paper, we use the term ``reflection" for the reflection
that takes place in the propagation path of the fast-mode
shock due to the non-uniform media. The ``echo" for the reflection
occurs on the boundary surface.
However, in physics, reflection and echo are actually the same phenomenon.

\begin{figure}
\centering
\includegraphics[width=9cm,clip,angle=0]{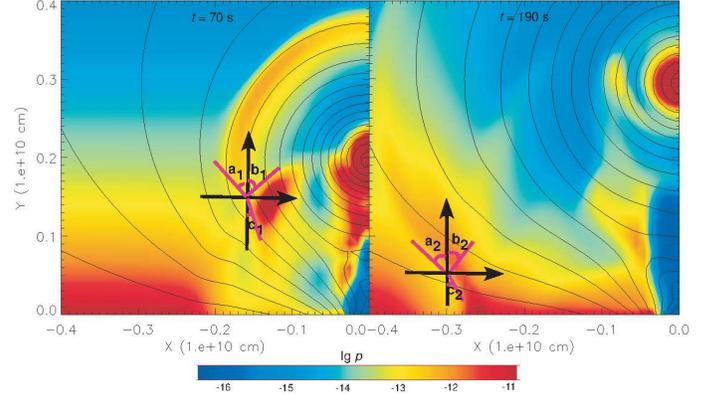}
\caption{Evolutions of the angle of incident, the angle of reflection and the angle of refraction
at two different time. $a_{1}$ and $a_{2}$ correspond to the angle of incident at $t = 70$~s and $t = 190$~s,
$b_{1}$ and $b_{2}$ are for the angle of reflection at $t = 70$~s and $t = 190$~s, $c_{1}$ and $c_{2}$
denote the angle of refraction at $t = 70$~s and $t = 190$~s.}
\label{fig:9}
\end{figure}

With the evolution progressing, two wake ends of the fast-mode shock sweep
side-back to the flank of the flux rope as shown in Fig.\,\ref{fig:8}.
The incident angle of the fast shock at the layer of its wake ends
becomes large. Panels in Fig.\,\ref{fig:8} clearly show that
the angle of the reflection and the angle of the refraction vary with
the incident angle. To further study this point, we investigate the variation of
the angle of the reflection and the angle of the refraction as the flux rope moves outward.
Figure \ref{fig:9} plots the evolutions of the angles of incidence, reflection
and refraction, respectively, at t = 70 s and t = 190 s.
With the downward-propagating fast shock encountering denser background plasma,
the incident angle becomes large, which results in
the increasing reflection and refraction angles between $t = 70$~s and $t = 190$~s.

\begin{figure}
\centering
\includegraphics[width=9cm,clip,angle=0]{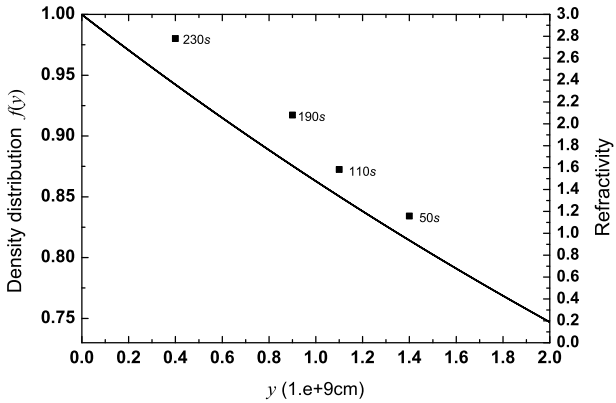}
\caption{The distribution of the background plasma density $f(y)$ and variation of
the fast shock refractivity versus height.
Solid points indicate the fast shock refractivity at different time. The curve represents $f(y)$ versus height.}
\label{fig:10}
\end{figure}

We also investigate the refractivity
of the fast shock at its wake ends and the distribution
of the background plasma density versus height as shown in Fig.\,\ref{fig:10}.
Solid points indicate the fast shock refractivity at different times.
The curve represents the background plasma density distribution versus
height, normalized at y=0. From Fig.\,\ref{fig:10}, we can see that the background plasma
density and the refractivity increase with decreasing height.

From the panels in Fig.\,\ref{fig:8}, we see that
the refraction of the fast-mode shock propagates
downward as the fast shock extends outward and sideward.
At about $t = 100$~s, the refracted wave touches the boundary
surface, and then reflection of the refracted wave occurs at
the boundary surface.
An echo of the refracted wave is thus produced at each of the
refracted wave's wake ends, which goes back into
the corona. This echo is weak, so it can not be easily distinguished.
At about $t = 260$~s, the fast-mode shock sweeps the boundary surface.
Because the background density gradient becomes sharp, the strong
reflection of the fast shock appears.

\begin{figure}
\centering
\includegraphics[width=9cm,clip,angle=0]{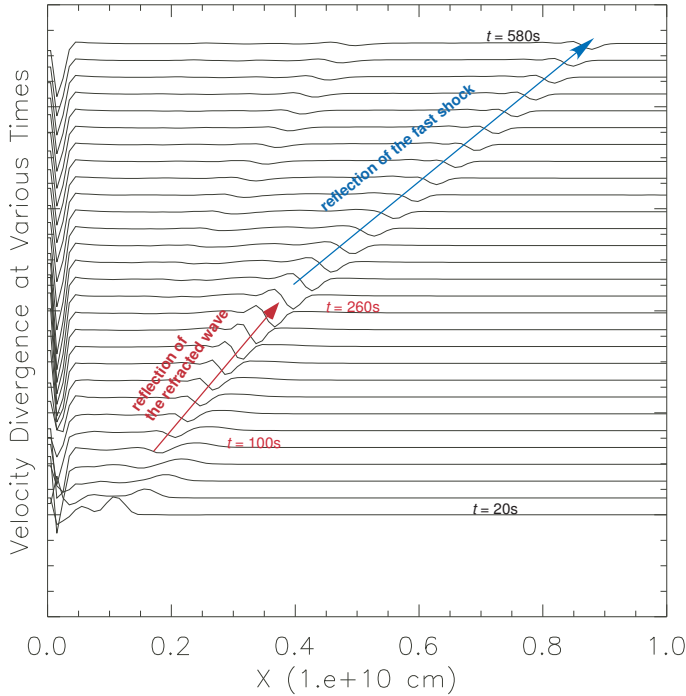}
\caption{Distributions of $\nabla\cdot \mathbf{v}$ on the boundary layer $y = 0.0$ in the time interval
between 20~s and 580~s.}
\label{fig:11}
\end{figure}

Figure \ref{fig:11} plots 29 distributions of $\nabla\cdot \mathbf{v}$
curves versus $x$ on the boundary $y = 0$ in the time interval between
$t = 20$~s and $t = 580$~s, which gives more information about propagation
of the refracted wave and the fast-mode shock at the boundary surface.
The plots clearly display two different
features of propagation at the boundary: the red arrow is for the footprint
of the refracted wave and the blue arrow is for the footprint of the fast-mode
shock. Figure \ref{fig:11} further confirms that the refracted wave approaches
the boundary surface at about 100~s. The fast shock starts to sweep the surface
at about 260~s as manifested in Fig.\,\ref{fig:8}.
By measuring the distances that they propagate in the given time interval,
we find that the speeds of the refracted wave and fast shock are 93 km s$^{-1}$
and 165 km s$^{-1}$, respectively, when they sweep the boundary
surface at y = 0.

As demonstrated in Figs.\,\ref{fig:6} and \ref{fig:8}, the fast
shock, its reflected wave and refracted wave (or reflection of the
refracted wave) interact with the vortices. At this stage, their
interaction could be related to the origin of the EUV waves.

\subsection{Formation of the Moreton wave and the secondary echo of the fast shock}

With the fast shock propagating outward and expanding downward,
the Moreton wave is invoked as the fast shock touches down at
about $t = 260$~s. As we illustrated in subsection 3.1, the fast
shock forms in front of the flux rope at about $t = 20$~s,
which then produces the type II burst \citep{Wang09,Lin06}.
This result indicates that the Moreton wave lags the type II burst
about 4 minutes, which is helpful to understand their correlation
observed in solar eruptions.

\begin{figure}
\centering
\includegraphics[width=9cm,clip,angle=0]{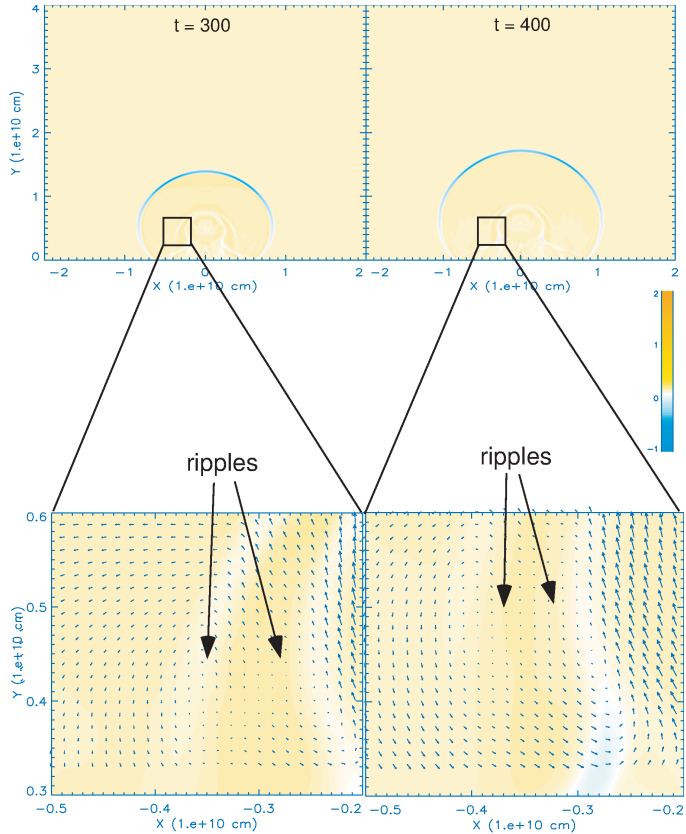}
\caption{Evolutions of $\nabla\cdot \mathbf{v}$ and streamlines at $t = 300$~s and $t = 400$~s.
The lower panels provide detailed structures in two subregions surrounded by rectangles marked in the upper panels,
respectively. The right color bar represents values of the velocity divergence in arbitrary unit.}
\label{fig:12}
\end{figure}

As the fast shock touches the boundary surface,
the echo of the fast shock will occur because the background density
gradient becomes sharp. The echo at the boundary mixes
with the reflection of the pre-existing wave, so the echo that
propagates back into the corona becomes so strong. It moves
sideward much faster than the initial reflected wave
(i.e. it appears on the layer above the boundary surface,
rather than the surface) of the fast shock.
Meanwhile, the initial reflected wave decays significantly
since the footpoint of the fast shock moves
outward and downward. On the other hand,
the velocity vortices propagate sideways and upward,
so the impact of the initial reflection of the fast-mode shock
on the vortices becomes weaker and weaker as illustrated
in Fig.\,\ref{fig:12}. In this figure, we plot the evolutions
of $\nabla\cdot \mathbf{v}$ and streamlines at
$300$~s and $400$~s. The lower panels provide detailed structures in
two subregions surrounded by the rectangles as marked in the upper panels.

The initial reflected wave of the fast-mode shock decays gradually and becomes weak.
However, the echo of the shock is fast and strong, producing one large
structure with two small vortices when it enters the region where the vortices
and the initial reflected wave are located. One of the two small vortices is
influenced by the initial reflected wave that appears at the same layer above the
boundary surface, while another one is impacted by the echo of the fast shock.
The latter is stronger and faster than the former, overtaking the former
and producing multiple ``ripples". \citet{Liu10} observed similar phenomena
in the 2010 April 8 event.

\begin{figure}
\centering
\includegraphics[width=9cm,clip,angle=0]{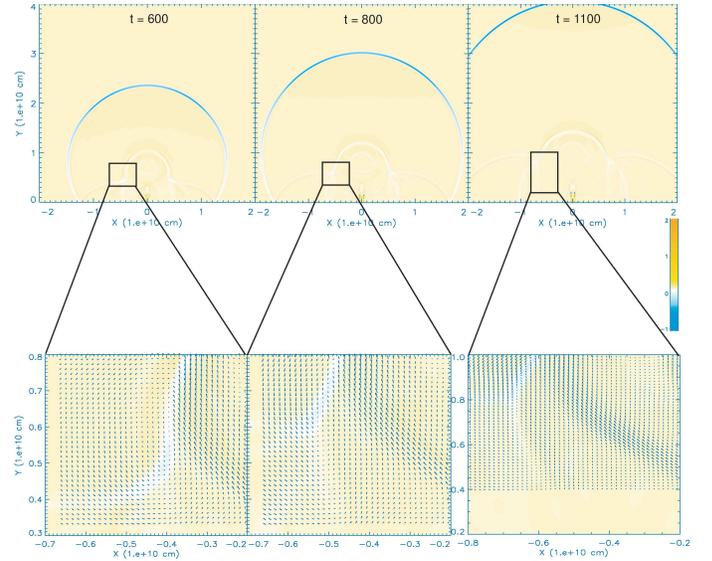}
\caption{Evolutions of $\nabla\cdot \mathbf{v}$ and streamlines at various times. The lower panels provide detailed structures
in three subregions surrounded by rectangles marked in the upper panels, respectively.
The right color bar represents values of the velocity divergence in arbitrary unit.}
\label{fig:13}
\end{figure}

\begin{figure}
\centering
\includegraphics[width=9cm,clip,angle=0]{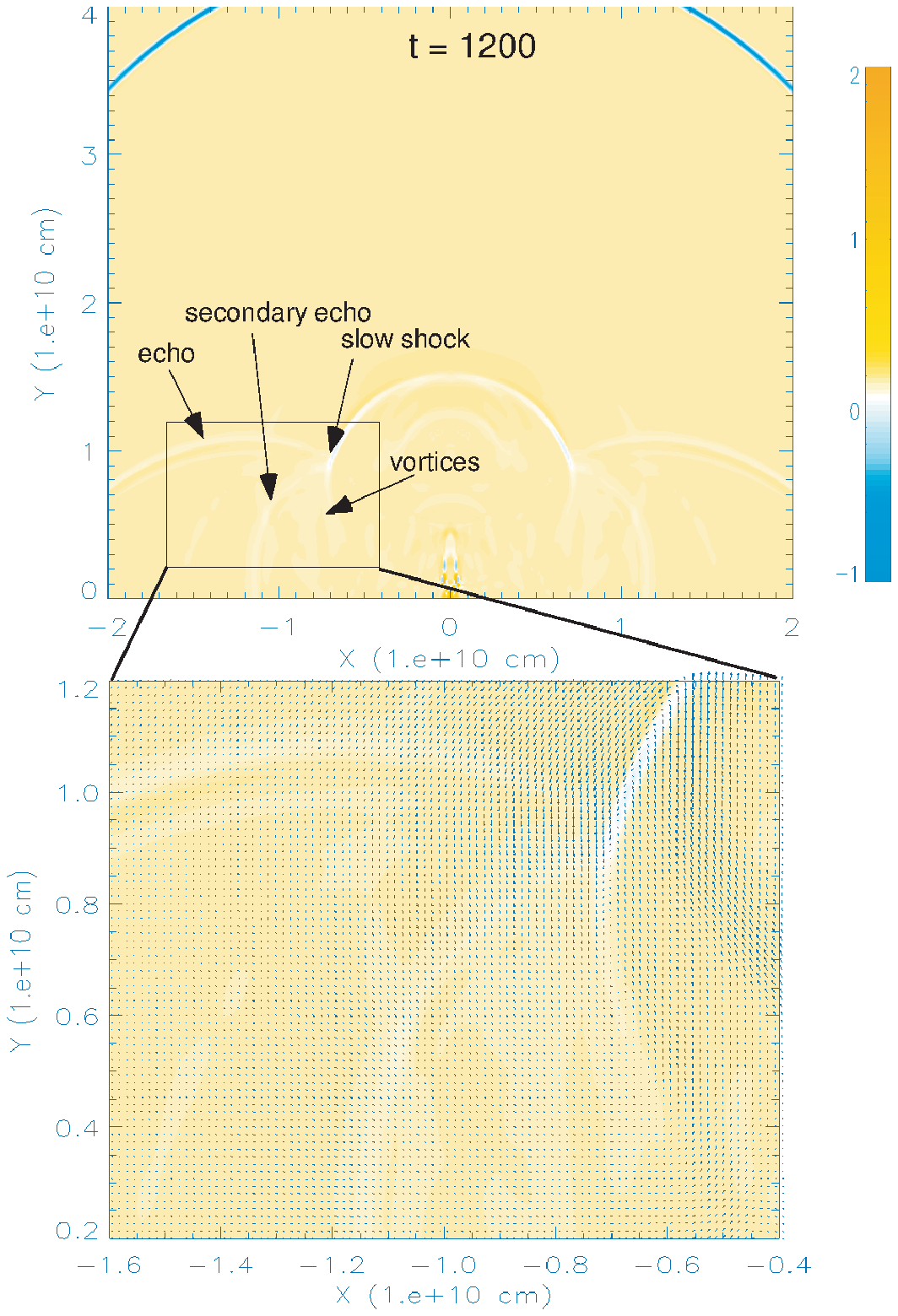}
\caption{Evolutions of $\nabla\cdot \mathbf{v}$ and streamlines at $t = 1200$~s. The echo, secondary echo, slow shock,
and vortices are denoted in the upper panel.
The lower panel shows detailed structure in a subregion surrounded by rectangle marked in the upper panel.}
\label{fig:14}
\end{figure}

With the flux rope moving up, the slow-mode shock is also generated
in front of the flux rope as shown in Fig.\,\ref{fig:13}. This figure
displays evolutions of $\nabla\cdot \mathbf{v}$ and streamlines at various
times. Unlike the fast-mode shock, the slow shock cannot reach the boundary
surface, because it starts to decay and dissipate above the surface \citep{Wang09,Mei12}.
From Figs.\,\ref{fig:13} and \ref{fig:14}, we see that the echo of the
fast-mode shock comes across the slow-mode shock before the slow shock
dissipates, and the reflection of the fast shock's echo (i.e. the secondary
echo of the fast shock)is excited in the region where the echo of the fast
shock encounters the slow shock as shown in Fig.\,\ref{fig:14}.

\begin{figure}
\centering
\includegraphics[width=9cm,clip,angle=0]{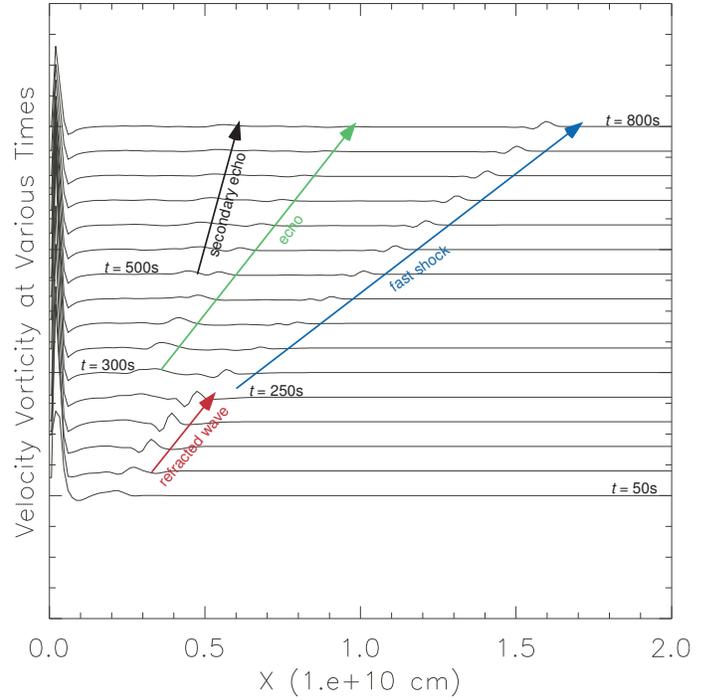}
\caption{Distributions of $(\nabla\times \mathbf{v})_{z}$ at the layer of $y = 3\times10^{4}$~km
that is in the corona. The time intervals for drawing these curves are between 50~s and 800~s.}
\label{fig:15}
\end{figure}

We also investigate the distributions of the $z$-component of
$\nabla\times\mathbf{v}$ at layer of $3\times10^{4}$~km from the
surface of the Sun as we did in the previous work \citep{Wang09}.
Figure \ref{fig:15} plots such distributions at various times.
The curve at the bottom is for $t = 50$ s, and the one at the top
is for $t = 800$~s. From Fig.\,\ref{fig:15}, we can see four propagating
signs: the blue arrow corresponds to the fast shock; the red arrow is for
the refracted wave of the fast shock; the green arrow shows the echo; the
black one denotes the secondary echo.

As we mentioned before, the inhomogeneity of the plasma causes
the fast shock to reflect and refract in the way as the
shock propagates towards the bottom boundary.
The shock reaches the layer of $3\times 10^{4}$~km at about $t=100$~s.
The reflection and the refraction of the shock take place simultaneously
at the same layer. At about $t=250$~s, the fast shock arrives at the
bottom boundary where the reflection becomes the echo and the refracted
wave disappears in our calculation. In fact, as indicated in Fig.\,\ref{fig:8},
the refracted wave should arrive at the bottom boundary
earlier than the fast-mode shock itself and produces an echo as well.
But this echo is weak and blended with other kinds of disturbances in the corona,
so we cannot recognize it in the present simulation.

After $t=250$~s, the fast shock sweeps
the boundary surface, and its echo propagates back to the corona.
At about $t = 300$ s, the echo gets to
the layer of $y=3\times 10^{4}$~km, which serves as an extra source of disturbance
in this layer as indicated in Fig.\,\ref{fig:15}.
The echo meets the slow mode shock at $t=500$~s, which gives rise to a further disturbance
in the region. Consistent with our previous discussions about
Fig.\,\ref{fig:7}, the disturbance directly caused by the fast shock
itself always moves faster than the others that are caused by various types of
secondary effect and their combinations.

By measuring the features seen in Fig.\,\ref{fig:15}, as we did in Fig.\,\ref{fig:11},
we find that the speed of the fast shock at layer $y=3\times 10^{4}$~km is 202~km~s$^{-1}$,
which is slightly slower than that at a lower layer (cf. Fig.\,\ref{fig:7}).
In addition, the speeds of the refracted wave, the echo and the second wave,
at the same layer, are 147 km s$^{-1}$, 123 km s$^{-1}$, and 50 km s$^{-1}$, respectively.
Furthermore, Fig.\,\ref{fig:15} depicts
a scenario of wave propagation in the corona where the EUV waves are usually
observed such that a fast component is followed by a group of slower components
with a large variety.

\begin{figure}
\centering
\includegraphics[width=9cm,clip,angle=0]{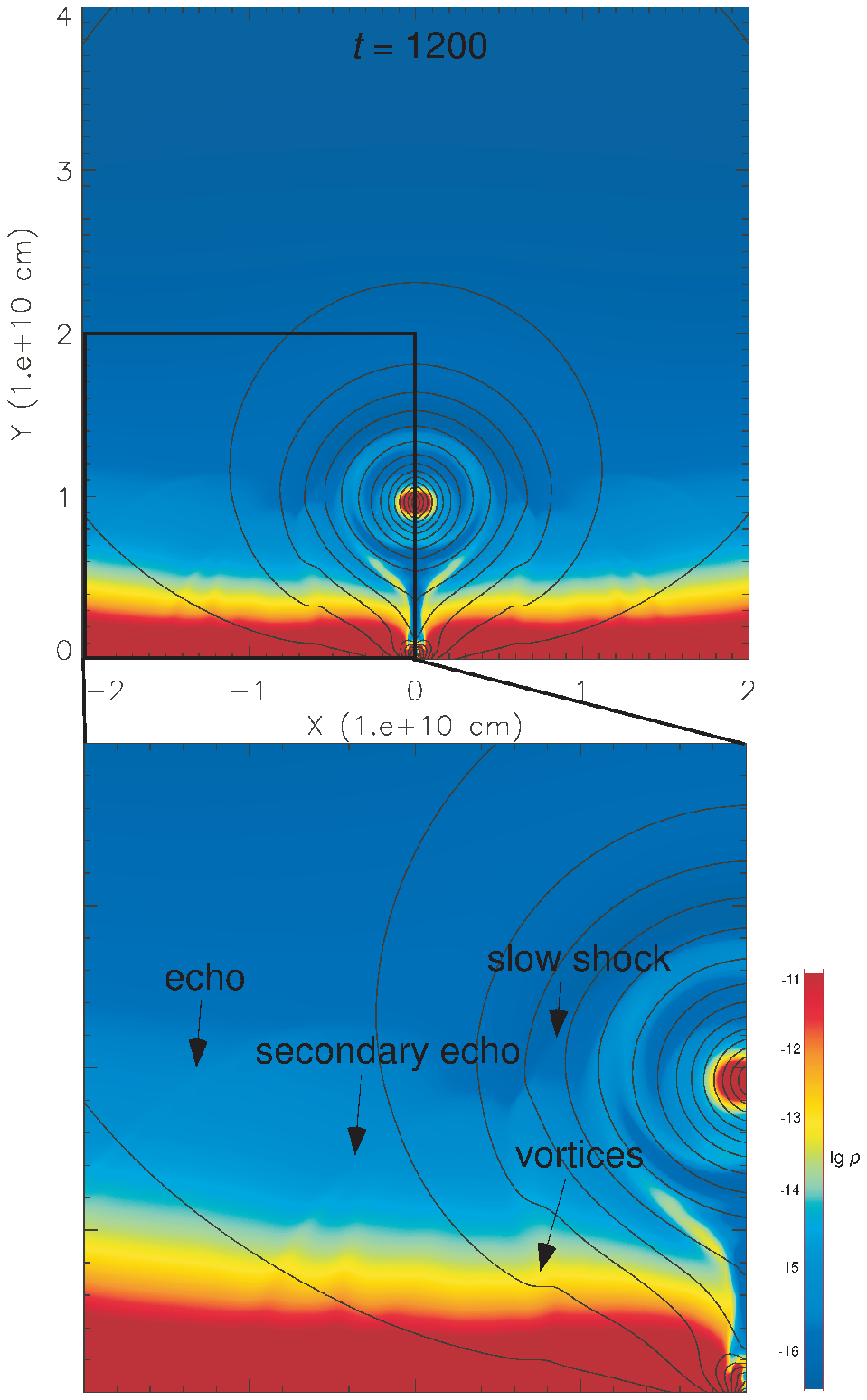}
\caption{Evolutions of magnetic field (black contours) and
the plasma density (shadings). The unit of time is second.
The right color bar represents values of the density in $\lg\rho$~(g~cm$^{-3})$.
A region surrounded by rectangle displays an area where the interaction between
the secondary echo propagating downward with the
velocity vortices results in distortion of the magnetic filed line.
Details in this region can be seen more clearly in the lower panel.}
\label{fig:16}
\end{figure}

The interaction between the secondary echo propagating downward with the
velocity vortices results in distortion of
the magnetic field line as indicated in Fig.\,\ref{fig:16},
which shows the evolutions of magnetic field and plasma density.
The continuous contours show magnetic field lines and the color shading the density distribution.
Figures \ref{fig:13}, \ref{fig:14} and \ref{fig:16} indicate
that the EUV waves in the last stage of propagation
may be contributed by combinations of the echo and the secondary echo
of the fast-mode shock, the velocity vortices, and the slow shock.

\section{Discussions}

Following our previous work \citep{Wang09}, we further studied
detailed evolutionary features of wave-like disturbances
and possible relation and/or contributions of these features
to the formation of the EUV waves.
In our study, the S\&G empirical atmosphere model is
employed for the density distribution of the background field,
and the present simulation has much higher resolution than that
of \citet{Wang09}.
With these improvement, we are able to perform more investigations
on the object of interest and to inspect more aspects of the relevant
issues. However, we need to note here that the purpose of our present
work is not to reproduce a specific EUV wave event as the works done by
\citet{Schmidt10} and \citet{Downs11,Downs12}.
Instead, our emphasis is to investigate how the solar
atmosphere would respond to an eruption that produces
CME and other associated activities, to explore what observational
consequences we could expect in such processes, and to see how these
consequences would contribute to the formation of EUV waves.

The most important discoveries or results of this work are
the reflection and refraction of the fast-mode shock during its
propagation, and formation of the secondary echo when
the first echo of the fast shock interacts with the slow mode shock
\citep{Wang09}. No numerical experiments have ever reported such a
phenomenon before.
The various disturbances that we identify in this work may
correspond to actual observational features related to the EUV wave
phenomenon observed in many solar eruptive processes.
Below we discuss our results one by one.

As the flux rope goes up, a fast-mode shock commences to form
forward of the flux rope at about $t = 20$~s, which then produced
the type II radio bursts \citep{Wang09,Lin06}.
Another important finding is that the velocity vortices begin
to separate from the flux rope when the fast shock forms.
These results duplicate what \citet{Cheng12} observed in a specific event.

Associated with the outward motion, expansion of the fast shock
takes place in various directions. The downward expansion shows the reflection and refraction
as a result of the non-uniform background plasma density.
In the uniform atmosphere \citep{Forbes90,Wang09},
the fast-mode shock propagates
through it without reflection and refraction.
In the non-uniform atmosphere, on the other hand,
reflection and refraction inevitably occur when
the fast shock reaches the boundary, separating
the media of different densities.
Because the density changes continuously in the real corona,
reflection and refraction take place everywhere in the
propagation path of the shock.
In our previous work \citep{Wang09,Mei12},
we did not notice this phenomenon because of the low
resolution of the grid used in those calculations.

The reflected component of the fast shock propagates upward, whereas the refracted
one propagates downward. As the refracted component
reaches the bottom boundary, an echo inevitably occurs.
But this echo may not be strong enough, so we cannot
distinguish its effect from that of other types of
disturbance in the same region (see Figs.\,\ref{fig:6}
and \ref{fig:8}). However, it surely contributes to the
formation of the slow component of EUV waves because
it is always located behind the fast-mode shock itself.
So our simulations show that the origin of EUV waves
includes not only the fast shock propagating freely in a
certain layer, but also various disturbances caused by
the fast shock and the motion of the flux rope.
This implies a complex origin of EUV waves.
This scenario is consistent with the results of \citet{Kwon13}.

With the fast shock expanding backward sequentially, the
background plasma becomes more dense. As a result, the incident
angle becomes large, increasing the reflection and refraction
angles. We also studied the refractivity of the fast shock versus
height, which shows that the background plasma density and
the refractivity increase with decreasing height.
At about $t = 100$~s, the refracted wave touches the boundary
surface and gets reflected, so an echo of the refracted wave
is created at each of the refracted wave's wake ends.
This echo is weak, so it is not easy to be distinguished.

When the fast shock sweeps the boundary surface, the strong
echo of the fast shock is produced because of the sharp density
gradient on the boundary. And then Moreton wave is invoked.
The Moreton wave lags the type II burst that is produced by the
formation of the fast shock about four minutes before,
which is helpful to understand their correlation observed in solar
eruptions.
Using the observations of the Hiraiso Radio Spectrograph in the
frequency range of 25-2500 MHz, \citet{Asai12} showed the observed
type II radio burst is connected to the Moreton wave and fast-bright
EUV wave. Furthermore, their results also support that the
Moreton wave originates from the fast shock.

More details of these features are revealed by the velocity divergence
$\nabla\cdot \mathbf{v}$ curves versus $x$ on $y = 0$,
which clearly displays two different features of propagation at the
boundary surface: one is the footprint of the refracted
wave; another is the footprint of the fast shock (Fig.\,\ref{fig:11}).
The result shows that the refracted wave speed at the boundary
is usually $1/2$ that of the fast shock at the boundary.

In addition to the fast-mode shock,
the slow-mode shock can also be recognized,
which is caused by the movement of the flux rope.
Unlike the fast shock, the slow shock begins to decay above the
boundary surface, so it cannot reach the surface.
Because the echo of the fast shock from the boundary surface
is strong, the secondary echo is excited when
the echo of the fast shock encounters the slow shock at some
layer above the boundary surface.
This phenomenon may have been observed in the events reported by
\citet{Shen14a} and \citet{Yang13}.

Furthermore, we point out that discussions
we made above revealed rich information about the origin of EUV waves.
We noticed that, except the disturbance caused directly
by the fast-mode shock itself, all the other types of disturbance
lag behind the fast shock, which has been confirmed by many observations
(e.g., \citet{Chen11b,Asai12,Shen12b,Shen12a,Kumar13,Shen14a}).
\citet{Chen02,Chen05} explained the EUV waves behind the fast-mode
shock as a result of the field line stretching by the CME motion,
and therefore the EUV wave is not a true ``wave", but a so-called pseudo
wave. Obviously, our results here, together with those of \citet{Wang09}
and \citet{Mei12}, provide an alternative explanation for the results
by \citet{Chen05}.

\section{Conclusions}

In this work, we use empirical atmosphere
S\&G model and a high resolution grid for numerical
simulations to study detailed evolutionary features
of wave-like disturbance.
Our main conclusions are summarized below.

1. Following the loss of equilibrium in the coronal magnetic structure,
the flux rope exhibits rapid motions, and invokes
the fast-mode shock in front of the rope, which then produces the type II radio bursts.

2. The velocity vortices form on either side of the flux rope,
and the vortices mix with the edge of the flux rope in the beginning.
However, the vortices start to separate from the edge of the flux rope
when the fast shock forms.

3. The fast shock expands downward and sideward simultaneously
as it propagates forward.
The downward expansion produces reflection and refraction
as a result of the non-uniform background plasma density.
The reflected component
of the fast shock propagates upward and the refracted component
propagates downward.

4. As the refracted component touches the boundary surface,
a weak echo is produced.
The Moreton wave is invoked as the fast shock sweeps the bottom boundary,
so the Moreton wave lags the type II burst.

5. As the fast shock touches the boundary surface, the strong echo
occurs because the background density gradient becomes sharp.
This echo propagates back into the corona.

6. Besides the fast-mode shock, the slow-mode shock is also excited
by the movement of the flux rope. Unlike the fast shock,
the slow shock begins to decay above the bottom boundary, so it cannot
reach the boundary. The secondary echo occurs in the area where the
echo of the fast shock encounters the slow shock. The nearby magnetic
field lines are further distorted because of the interaction between
the secondary echo and the vortices.

7. During the subsequent evolution, we found an interesting phenomenon:
a large structure consisting of two small vortices. One vortex is influenced
by the initial reflected component of the fast shock, and another one is impacted
by the strong echo of the fast shock. The latter vortex is stronger and faster,
and so it overtakes the former in a short time scale, which may correspond to
the ripples as observed in a specific event (e.g. see \citet{Liu10}).

Our results indicate that various origins of the EUV waves exist.
These include the fast shock,
the reflected and refracted component of the fast shock, the echo of the
refracted component, the echo and the secondary echo of the fast shock,
the slow shock, and the velocity vortices (e.g., \citet{Forbes90,Wang09,Mei12}).
Considering that the EUV waves appear roughly in the same layer
as the disturbance that is usually left behind the fast-mode shock,
we conclude that the slow component of the EUV waves may be invoked by
one of (or the combination of) these sources, and the fast EUV wave component
should be due to the fast-mode shock itself.

We are grateful to the referee for his/her valuable and constructive comments
and suggestions. We also thank our colleague Y. Shen for the discussions.
This work was supported by the National Basic Research Program of China
(2012CB825600 and 2011CB811406), and the Shandong Province Natural Science
Foundation (ZR2012AQ016). JL's work was supported by the Program 973 grants
2011CB811403 and 2013CBA01503, the NSFC grants 11273055 and 11333007,
and the CAS grant XDB09040202.

\bibliographystyle{aas}
\bibliography{apjref}

\end{document}